\documentstyle[12pt]{article}

\advance \topmargin by -\headheight
\advance \topmargin by -\headsep
     
\evensidemargin \oddsidemargin
\begin{document}
 
\topmargin -.6in

%               macros formatting and equations
\def\rf#1{(\ref{eq:#1})}
\def\lab#1{\label{eq:#1}} 
\def\br{\begin{eqnarray}}
\def\er{\end{eqnarray}}
\def\be{\begin{equation}}
\def\ee{\end{equation}}
\def\nonu{\nonumber}
\def\lb{\lbrack}
\def\rb{\rbrack}
\def\({\left(}
\def\){\right)}
\def\v{\vert}
\def\bv{\bigm\vert}
\def\lskip{\vskip\baselineskip\vskip-\parskip\noindent}
\relax
\newcommand{\nit}{\noindent}
\newcommand{\ct}[1]{\cite{#1}}
\newcommand{\bi}[1]{\bibitem{#1}}
%%                                math symbols
%
\def\a{\alpha}
\def\b{\beta}
\def\ca{{\cal A}}
\def\cm{{\cal M}}
\def\cn{{\cal N}}
\def\cf{{\cal F}}
\def\d{\delta}
\def\D{\Delta}
\def\eps{\epsilon}
\def\g{\gamma}
\def\G{\Gamma}
\def\grad{\nabla}
\def\h{ {1\over 2}  }
\def\hc{\hat{c}}
\def\hd{\hat{d}}
\def\hg{\hat{g}}
\def\hp{ {+{1\over 2}}  }
\def\hm{ {-{1\over 2}}  }
\def\k{\kappa}
\def\l{\lambda}
\def\L{\Lambda}
\def\lg{\langle}
\def\m{\mu}
\def\n{\nu}
\def\o{\over}
\def\O{\Omega}
\def\p{\phi}
\def\pa{\partial}
\def\pr{\prime}
\def\ra{\rightarrow}
\def\rh{\rho}
\def\rg{\rangle}
\def\s{\sigma}
\def\t{\tau}
\def\th{\theta}
\def\ti{\tilde}
\def\wti{\widetilde}
\def\inte{\int dx }
\def\xb{\bar{x}}
\def\yb{\bar{y}}
%%                     common physics symbols
\def\tr{\mathop{\rm tr}}
\def\Tr{\mathop{\rm Tr}}
\def\partder#1#2{{\partial #1\over\partial #2}}
\def\ds{{\cal D}_s}
\def\wtwo{{\wti W}_2}
%%                    macros for Lie algebras
\def\lie{{\cal G}}
\def\alie{{\widehat \lie}}
\def\dlie{{\cal G}^{\ast}}
\def\elie{{\widetilde \lie}}
\def\edlie{{\elie}^{\ast}}
\def\hlie{{\cal H}}
\def\wlie{{\widetilde \lie}}
%%       fake blackboard bold macros for reals, complex, etc.
\def\rlx{\relax\leavevmode}
\def\inbar{\vrule height1.5ex width.4pt depth0pt}
\def\IZ{\rlx\hbox{\sf Z\kern-.4em Z}}
\def\IR{\rlx\hbox{\rm I\kern-.18em R}}
\def\IC{\rlx\hbox{\,$\inbar\kern-.3em{\rm C}$}}
\def\one{\hbox{{1}\kern-.25em\hbox{l}}}

%       This defines the journal citations
%
\def\PRL#1#2#3{{\sl Phys. Rev. Lett.} {\bf#1} (#2) #3}
\def\NPB#1#2#3{{\sl Nucl. Phys.} {\bf B#1} (#2) #3}
\def\NPBFS#1#2#3#4{{\sl Nucl. Phys.} {\bf B#2} [FS#1] (#3) #4}
\def\CMP#1#2#3{{\sl Commun. Math. Phys.} {\bf #1} (#2) #3}
\def\PRD#1#2#3{{\sl Phys. Rev.} {\bf D#1} (#2) #3}
\def\PLA#1#2#3{{\sl Phys. Lett.} {\bf #1A} (#2) #3}
\def\PLB#1#2#3{{\sl Phys. Lett.} {\bf #1B} (#2) #3}
\def\JMP#1#2#3{{\sl J. Math. Phys.} {\bf #1} (#2) #3}
\def\PTP#1#2#3{{\sl Prog. Theor. Phys.} {\bf #1} (#2) #3}
\def\SPTP#1#2#3{{\sl Suppl. Prog. Theor. Phys.} {\bf #1} (#2) #3}
\def\AoP#1#2#3{{\sl Ann. of Phys.} {\bf #1} (#2) #3}
\def\PNAS#1#2#3{{\sl Proc. Natl. Acad. Sci. USA} {\bf #1} (#2) #3}
\def\RMP#1#2#3{{\sl Rev. Mod. Phys.} {\bf #1} (#2) #3}
\def\PR#1#2#3{{\sl Phys. Reports} {\bf #1} (#2) #3}
\def\AoM#1#2#3{{\sl Ann. of Math.} {\bf #1} (#2) #3}
\def\UMN#1#2#3{{\sl Usp. Mat. Nauk} {\bf #1} (#2) #3}
\def\FAP#1#2#3{{\sl Funkt. Anal. Prilozheniya} {\bf #1} (#2) #3}
\def\FAaIA#1#2#3{{\sl Functional Analysis and Its Application} {\bf #1} (#2)
#3}
\def\BAMS#1#2#3{{\sl Bull. Am. Math. Soc.} {\bf #1} (#2) #3}
\def\TAMS#1#2#3{{\sl Trans. Am. Math. Soc.} {\bf #1} (#2) #3}
\def\InvM#1#2#3{{\sl Invent. Math.} {\bf #1} (#2) #3}
\def\LMP#1#2#3{{\sl Letters in Math. Phys.} {\bf #1} (#2) #3}
\def\IJMPA#1#2#3{{\sl Int. J. Mod. Phys.} {\bf A#1} (#2) #3}
\def\AdM#1#2#3{{\sl Advances in Math.} {\bf #1} (#2) #3}
\def\RMaP#1#2#3{{\sl Reports on Math. Phys.} {\bf #1} (#2) #3}
\def\IJM#1#2#3{{\sl Ill. J. Math.} {\bf #1} (#2) #3}
\def\APP#1#2#3{{\sl Acta Phys. Polon.} {\bf #1} (#2) #3}
\def\TMP#1#2#3{{\sl Theor. Mat. Phys.} {\bf #1} (#2) #3}
\def\JPA#1#2#3{{\sl J. Physics} {\bf A#1} (#2) #3}
\def\JSM#1#2#3{{\sl J. Soviet Math.} {\bf #1} (#2) #3}
\def\MPLA#1#2#3{{\sl Mod. Phys. Lett.} {\bf A#1} (#2) #3}
\def\JETP#1#2#3{{\sl Sov. Phys. JETP} {\bf #1} (#2) #3}
\def\JETPL#1#2#3{{\sl  Sov. Phys. JETP Lett.} {\bf #1} (#2) #3}
\def\PHSA#1#2#3{{\sl Physica} {\bf A#1} (#2) #3}
\def\PHSD#1#2#3{{\sl Physica} {\bf D#1} (#2) #3}
%%%

\begin{titlepage}
\vspace*{-2 cm}
\noindent
%\hfill{hep-th/9803122 }\\

\begin{flushright}
IFT--P.017/97\\ 
CBPF--NF--018/98 
\end{flushright}

\vskip3cm

\begin{center}
{\Large\bf $SU(2,R)_q $ Symmetries of Non-Abelian Toda Theories }
\vglue 2  true cm
{ J. F. Gomes}\footnote{Partially 
supported by a CNPq research grant}$^{,\dagger}$,
 { G. M. Sotkov}\footnote{On leave of absence from the Institute for Nuclear
 Research and Nuclear Energy, Bulgarian Academy of Sciences, 1784, Sofia,
 Bulgaria}$^*$ and { A. H. Zimerman}$^{1,\dagger}$\\

\vspace{1 cm}

$^{\dagger}${\footnotesize Instituto de F\'\i sica Te\'orica - IFT/UNESP\\
Rua Pamplona 145\\
01405-900, S\~ao Paulo - SP, Brazil}\\
jfg@axp.ift.unesp.br, zimerman@axp.ift.unesp.br\\

\vspace{1 cm} 

$^*${\footnotesize Centro Brasileiro de Pesquisas F\'\i sicas-CBPF\\
Rua Xavier Sigaud 150\\
22290-180, Rio de Janeiro, RJ}\\
sotkov@cbpfsu1.cat.cbpf.br\\
\medskip
\end{center}

\normalsize
\vskip3cm

\begin{center}
{\large {\bf ABSTRACT}}\\
\end{center}
%\par \vskip .1in

%\vskip .5 cm

\noindent
{\footnotesize 
The classical and quantum algebras of a class of conformal 
 NA-Toda models are studied.It is shown that  the $SL(2,R)_q$ Poisson brackets algebra
 generated  by certain chiral and antichiral charges of the nonlocal currents
 and the global $U(1)$ charge appears as an algebra of the 
 symmetries of these models.  } 
\vglue 1 true cm

\end{titlepage}

%%%%%%%%%%%%%%%%%%%%%%%%%%%%%%%%%%%%%%%%%%%%%%%%%%%%%%%%%%%%%%%%%%%%%%%%%%%%%%

%\section{Introduction}
The 
(Non Abelian) NA-Toda theories are singled out among all 2-D conformal 
models (CFT's)
 by their
important role in the construction of exact solutions for both self-dual 4-D
Yang-Mills theories (axial symmetric instantons) 
\cite{Witten1,Bais,Leznov}
and string theory (black hole backgrounds)\cite{Witten2,Dijkgraaf,Gervais}.
 One  expects that their
quantum counterparts will provide an appropriate statistical mechanical 
tools for describing the critical behaviour of   the $SU(n)$ gauge theory. 
  The progress in the quantization of large class of 2-D
 CFT's based
on the Abelian Toda models and their algebra of symmetries - $W_n$ mininal
 models,
$W_n$-gravities, etc, \cite{Fateev1,Fateev2,Bilal1,Bershadsky1,Knizhnik,Belavin} suggests that similar algebraic strategy of
 quantization takes
place for the NA-Toda as well.  Few important steps
 in this direction
concerning  
the construction
of conserved currents and their ({\it non-local and non-Lie}) classical
(Poisson brackets ) algebras for the
 particular  case of $B_2$
NA-Toda \cite{Gervais,Bilal2}
 have been realized. 
 
This letter  is devoted to the construction of classical and quantum algebras
 of symmetries of the first few members ($n=1,2,3$) of the following family of
 $A_n$-NA-Toda models: 
\br
L & =& -{k\o {2 \pi}} \{{1\o 2} \eta_{ik} \pa \phi_i \bar \pa \phi_k 
 -({2\o k})^2 \sum_{i=1}^{n-1}
e^{\phi_{i-1} +\phi_{i+1} -2 \phi_i}) +
{1\o {2\Delta }}e^{-\phi_1} (\pa \psi \bar \pa \chi +
 \pa \chi \bar \pa \psi )-
\nonumber \\
& -&  {1\o {4\Delta }}e^{-\phi_1}[\bar \pa \phi_1 (\chi \pa \psi - \psi \pa \chi )
  - \pa \phi_1
 (\chi \bar \pa \psi - \psi \bar \pa \chi )] \}  
\label{1}
\er
where $\eta_{ik} =2\d_{i,k}-\d_{i,k+1} - \d_{i,k-1}$, $\Delta = 
 1+ {n+1 \o 2n} e^{-\phi_1} \psi \chi $ and 
$ \pa = \pa_{\tau} + \pa _{\sigma}$, $ \bar \pa = \pa_{\tau} - 
\pa _{\sigma} $,
 $\phi_0 = \phi_n =0$.  They represent the (non-compact)
 $SL(2,R)/U(1)$-parafermions interacting with the $A_{n-1}$-abelian Toda model. 
 One can derive (\ref{1}) as an effective Lagrangean for the gauged $H_-
 \setminus A_n /H_+$-WZW model \footnote{see for more details our forthcoming
 paper \cite{GSZ}}, $H_{\pm} = N_{(1)}^{\pm} \otimes H_0^{0 (1)}$ where 
 $N_{(1)}^{\pm}$ are nilpotent subgroups of $A_n$ spanned by $E_{[\a ]_1}$ or 
$E_{-[\a ]_1}$   (${[\a ]_1}$- all positive roots but $\a_1 $) and 
 $H_0^{0 (1) }= exp \{{R(z, \bar z)\lambda_1 H}\}$.  This is equivalent to
 consider a specific Hamiltonian reduction of the $A_n$-WZW model by imposing
 the following set of constraints:
 \br
&&J_{-\a_i} = \bar J_{\a_i}= 1,\, i=2, \ldots , n ; \,\,\,
J_{\lambda_1H} =\bar J_{\lambda_1H} = 0
\nonumber \\ 
&&J_{-[\a ]} = \bar J_{[\a ]}= 0, \,\,\,{\rm for}\,\, \a \; \;
{\rm nonsimple\, \, positive\,\,  roots}
\label{2}
\er
(i.e. $J_{-\a_1}$ and   $\bar J_{\a_1}$ are left {\it unconstrained}).  Taking
into account the residual gauge symmetries (that keep (\ref{2}) unchanged ) we
find the reduced form of the conserved currents:
\br
J &=& V^+E_{-\a_1} + \sum_{i=2}^{n}E_{-\a_i} + V^-E_{\a_1+ \cdots +\a_n} +
\sum_{i=2}^{n}W_{n-i+2}E_{\a_i + \a_{i+1} + \cdots + \a_n} \nonumber \\
\bar J &=& \bar V^+E_{\a_1} + \sum_{i=2}^{n}E_{\a_i} +
 \bar V^-E_{-\a_1- \cdots -\a_n} +
\sum_{i=2}^{n}\bar W_{n-i+2}E_{-\a_i - \a_{i+1} - \cdots - \a_n}
\label{3}
\er
The well known fact is that the remaining currents $V^{\pm}$, $W_{n-i+2}$ (
and $\bar V^{\pm}$, $\bar W_{n-i+2}$) appears as conserved currents of the
reduced model (\ref{1}).  The spins of these currents calculated with respect
to the improved stress tensor ($T=W_2$):
$$
T^{imp} = {1\o {k+n+1}} Tr :{J^a J^a }: +
 \sum_{i=2}^{n} \lambda_i\cdot H \pa J_i
- {(n-1)\o 2} \lambda_1\cdot H \pa J_1
\nonumber 
$$
are given by $s^{\pm} =s(V^{\pm}) = {(n+1)\o 2}$, $s_i=s(W_{n-i+2})= n-i+2$. 
Note that our Lagrangean (\ref{1}) is invariant under a global $U(1)$ gauge
transformation : $\psi^{\pr} = e^{\a } \psi$, $ \chi^{\pr} = e^{-\a } \chi$
and $\phi_i^{\pr} = \phi_i $ ($\a$ is a constant).  The corresponding {\it 
nonchiral } $U(1)$-current :
\be
  J_{\mu} = -{{k \o {4\pi}}}  ( \chi \pa_\mu
\psi - \psi \pa_{\mu} \chi + \psi \chi\pa _{\mu} \phi_1  ) 
{e^{-\phi_1}\o {\Delta }}
 \label{4}
 \ee
($ \pa \bar J + \bar \pa J = 0 $ where $ J = {1 \o 2} (J_0 + J_1 )$,
 $\bar  J = {1 \o 2} (J_0 - J_1)$) completes the list of the conserved currents
 of the NA-Toda models given by (\ref{1}).  Starting from the WZW currents $J =
 {k\o 2}g^{-1}\pa g$, $\bar J = -{k\o 2} \bar \pa g g^{-1}$ ( $ g=
 e^{\chi_{[\a ]}E_{-[a ]}} e^{\phi_i \lambda_iH} e^{\psi_{[\a ]}E_{[\a ]}}$)
 and solving the constraints  and the gauge fixing conditions in terms of the
 physical fields $\psi = \psi_1 e^{-{{n+1}\o 2n}R},
  \chi = \chi_1 e^{-{{n+1}\o 2n}R}$ and $\phi_i, (i=2, 3, \cdots ,n$) only one
  can {\it in principle } derive the explicit form of the remaining currents. 
  We find that, for example, $V^{+}, \bar V^{-}$   and $T (\bar T)$ are given
  by :
$$
V^+ = {k\o 2} {{e^{-\phi_1 + {{n+1 \o 2n}R} }\pa
\chi }\o {\Delta }}
\quad \quad \quad 
\bar V^- = {k\o 2} {{e^{-\phi_1 + {{n+1 \o 2n}R}}    
\bar \pa \psi }\o {\Delta } }
\nonumber 
$$
\be
T(z) = {1\o 2}\eta_{ik}\pa \phi_i \pa \phi_k 
 +\sum_{i=1}^{n-1}\pa^2 \phi_i + \pa \chi \pa \psi
{e^{-\phi_1} \o {\Delta }}  + {{n-1}\o 2}\pa ({{\psi \pa \chi }\o \Delta
}e^{-\phi_1})
\label{5}
\ee 
and $\bar T= T(\pa , \psi , \chi \longrightarrow \bar \pa , \chi , \psi )$.  The
$R(z, \bar z )$ is the nonlocal field that denotes the solution of the
following system of equations :
\be
\pa R = {e^{-\phi_1} \o \Delta }\psi \pa \chi  
 \quad \quad \quad ,
\bar \pa R = {e^{-\phi_1}\o \Delta }\chi \bar \pa \psi 
\label{6}
\ee
which are nothing but the constraint equations  $J_{\lambda_1H} = 
\bar J_{\lambda_1H} =0$.  One can eliminate $R$ from (\ref{5}) by solving eqn
(\ref{6}) and thus introducing certain {\it nonlocal } terms in $V^{\pm}$ and
$\bar V^{\pm }$.  The {\it nonlocality } of the $V^{\pm}$ and $\bar V^{\pm}$ is
one of the {\it main features } of the NA-Toda models (\ref{1}) originated from
the additional {\it PF-type constraint }$J_{\lambda_1H}=\bar J_{\lambda_1H} = 0$.
Note that all the others  $W_{n-i+2}$-currents are {\it local }.  To derive
their explicit form, as well as the form of $V^-$ and $\bar V^+$ for arbitrary
$n$ is quite a difficult problem.  For $n=1$ we obtain:
\be 
V_{n=1}^-(z) = {k \o 2} e^{-R} \pa \psi \;\;, \;\;
 \bar V_{n=1}^+(z) = {k \o 2} e^{-R}
\bar \pa \chi
\label{7}
\ee
(both of spin 1) and for $n=2$ the remaining nonlocal current (of spin ${3\o
2}$) are given by
\be
 V_{n=2}^-(z) = ( {k \o 2})^2  e^{-{3\o 4}R} \(\pa^2 \psi +
 {1\o 16}\psi (\pa R)^2
- \psi(\pa \phi_1)^2-\psi\pa^2 \phi_1 - 
{1 \o 4} \psi \pa^2 R - {1 \o 2} \pa \psi  \pa R \)
\label{8}
\ee
 and 
$ \bar V^+_{n=2} = V_{n=2}^-$ $(\psi \rightarrow \chi$,
 $\pa \rightarrow \bar \pa)$. 
 
 The simplest method \cite {Polyakov}  for deriving the chiral algebra of symmetries of (\ref{1})
 spanned by $V^{\pm}, W_{n-i+2}$, ($i=2,3, \cdots ,n$) consists in imposing the
 constraints (\ref{2}) and the gauge fixing conditions directly in the
 $A_n$-gauge transformations: $\d J = [\epsilon ,J] -{k\o 2} \pa \epsilon $. 
 This leads to the following system of first order differential equations 
 \be
 -{k\o 2} \pa \epsilon_{ik} = J_{ij}\epsilon_{jk} - \epsilon_{ij}J_{jk}, \quad
 \quad (ik) \neq \{ (2,1), (p, n+1) (1,n+1)\}
 \label{9}
 \ee
 for $i,j,k = 1,2, \cdots , n+1$.  
 The equations for $(i,k) = \{(2,1), (p,n+1), (1,n+1)\}$
$$
\d J_{ik} = \eps_{ij}J_{jk} - J_{ij}\eps_{jk} - {k\o 2} \pa \eps_{ik}
\nonumber
$$
($J_{12}=V^{+}, J_{1,n+1}=V^-, J_{p,n+1} =W_{n-p+2}, p=2,3,\cdots ,n-1$) gives
the transformation law for the remaining currents we are looking for.  The
problem is to solve (\ref{9}) for the redundant $\eps_{ik}$'s in terms of the
independent parameters $\eps_{12}=\eps^-, \eps_{n+1,1} = \eps^+,
\eps_{n+1,n}=\eps , \eta_{n-p+2}=\eps_{n+1,p}, p=2,3, \cdots ,n-1$ and the
currents $V^{\pm}, W_{n-i+2}$.  The transformations generated by the nonlocal
currents $V^{\pm}$ are given by ($\tilde \eps^{\pm} = -{k\o 2}\eps^{\pm}$):
\br
\d_{\eps^{\pm}}V^- &=& {{n+1}\o {nk^2}} 
\int \eps (\sigma - \sigma^{\pr})[\tilde
\eps^+(\sigma^{\pr})V^-(\sigma^{\pr}) - \tilde \eps^-(\sigma^{\pr})V^+(\sigma^{\pr})]
V^{-}(\sigma )d\sigma^{\pr} \nonumber \\
&-&\sum_{s=0}^{n-2}({k\o 2})^{s-1}W_{n-s}\pa ^{s}\tilde \eps^-(\sigma ) + 
({k\o 2})^{n-1}\pa ^{n}\tilde \eps^- (\sigma )
\label{10}
\er
and similar for $\d_{\eps^{\pm}} V^{+}$.  Reminding the relation between the
infinitesimal transformations and the currents Poisson brackets:
$$
\d_{\eps^{\pm}} I(\sigma ) = \int d\sigma^{\pr}  \eps^{\pm}(\sigma^{\pr}
)\{V^{\mp}(\sigma^{\pr} ), I(\sigma ) \}
\nonumber 
$$
$(I = V^+$ or $V^-)$ we deduce from (\ref{10}) the form of the algebra of the
nonlocal currents $V^{\pm}$:
\br
&&\{ V^{\pm} (\sigma),   V^{\pm} (\sigma^{\pr})\}
 = - {{n+1} \o nk^2 } \epsilon (\sigma - \sigma^{\pr})V^{\pm}
(\sigma)V^{\pm}(\sigma^{\pr})
\nonumber \\
&&\{ V^{+} (\sigma),   V^{-} (\sigma^{\pr}) \}
  =  {{n+1} \o nk^2 } \epsilon (\sigma - \sigma^{\pr})V^{+}
(\sigma)V^{-}(\sigma^{\pr})+ ({k\o 2})^{n-1} \pa_{\sigma^{\pr}}^{n} \d (\sigma -
\sigma^{\pr})   \nonumber \\
 &&\hskip 3cm  -   \sum_{s=0}^{n-2}({k\o 2})^{s-1} W_{n-s}(\sigma ^{\pr})
  \pa ^{s}_{\sigma^{\pr}}
\d (\sigma - \sigma^{\pr})
\label{11}
\er
Leaving the general solution of (\ref{9}) to our forthcoming paper \cite {GSZ},
we consider here a few particular cases $n=1,2,3$. The algebra of the
symmetries $V_2^{(1,1)}$ of  the $A_1$-NA-Toda model is generated by
$V^{\pm}_{n=1}$ ( of spin 1 ) only (i.e. eqn. (\ref{11}) for $n=1$).  It is
identical to the semiclassical limit of the PF-algebra \cite{Zamolodchikov} studied
in ref. \cite{Bardakci}.  In the $n=2$ case the corresponding $V_3^{(1,1)}$
algebra contains appart from (\ref{11}), which now reads,
\br
\{ V^{+} (\sigma),   V^{-} (\sigma^{\pr}) \}
  =  {{3} \o {2k^2} } \epsilon (\sigma - \sigma^{\pr})V^{+}
(\sigma)V^{-}(\sigma^{\pr})  
%\nonumber \\
 -  {2\o k}T(\sigma^{\pr})\d (\sigma - \sigma^{\pr})  + 
 {2\o k}\pa _{\sigma^{\pr}}
\d (\sigma - \sigma^{\pr})
\nonumber
\er
($T=W_2$) the PB's of $T$ and $V^{\pm}$
\be
\{T(\sigma), V^{\pm}(\sigma^{\pr})\} =
sV^{\pm}(\sigma^{\pr})\pa_{\sigma^{\pr}} \d(\sigma - \sigma^{\pr})+
\d (\sigma - \sigma^{\pr}) \pa_{\sigma^{\pr}}V^{\pm}(\sigma^{\pr}) 
\label{12}
\ee 
with $s= {3\o 2}$ and the usual Virasoro subalgebra 
\be
\{T(\sigma), T(\sigma^{\pr})\} = 2T( \sigma^{\pr}) 
\pa_{\sigma^{\pr}}\d (\sigma -\sigma^{\pr}) + 
\pa_{\sigma^{\pr}}T(\sigma^{\pr})\d (\sigma - \sigma^{\pr}) 
  -{k^2 \o 2} \pa ^3_{\sigma^{\pr}}\d (\sigma - \sigma^{\pr}). 
\label{13}
\ee
The $n=3$ case appears to be a {\it nonlocal } and {\it nonlinear }( quadratic terms
) extension of the Virasoro algebra (\ref{13}) with two spin $s=2$ nonlocal 
currents $V^{\pm}_{n=3}$ and one local  $W_3$ of spin $s=3$.  The
$V_4^{(1,1)}$-algebra combines together the features of the PF-algebra  and the
$W_3$-one.  Apart from the (\ref{11}), (\ref{12}) and (\ref{13}) ( with central
charge $-2k^2$) we have two new PB's
\br
&&\hskip-1.5cm\{w_3(\sigma)  , V^{\pm}(\sigma^{\pr})\} =  \mp {5k \o 3 } 
(\pa _{\sigma ^{\pr}} ^2 \d
(\sigma -
\sigma^{\pr})) V^{\pm} (\sigma^{\pr}) \mp {5k \o 2}
 ( \pa_{\sigma ^{\pr}} \d (\sigma -
\sigma^{\pr}))
\pa _{\sigma^{\pr}}V^{\pm}(\sigma^{\pr})     \nonumber \\
&&\hskip1.5cm \pm   \d (\sigma - \sigma^{\pr}) ({2 \o {3k}}T
V^{\pm} - k
\pa _{\sigma^{\pr}}^2 V^{\pm}) 
\nonumber \\
&&\hskip-2cm\{ w_3(\sigma) , w_3(\sigma^{\pr}) \}   =  4(\pa _{\sigma ^{\pr}} \d (
\sigma -\sigma^{\pr}  ))(V^+ V^- + {1 \o 6} T^2) (\sigma^{\pr})  %\nonumber \\
+  2 \d (\sigma -
\sigma^{\pr})
\pa _{\sigma^{\pr}} (V^+V^- + {1\o 6} T^2) (\sigma^{\pr})
\nonumber \\
&&\hskip0.5cm 
-  {3k^2 \o 4} ( \pa_{\sigma^{\pr}}\d(\sigma -
\sigma^{\pr}))
\pa^2_{\sigma^{\pr}} T(\sigma^{\pr})    - 
{5k^2 \o 4} ( \pa^2_{\sigma^{\pr}}\d (\sigma -
\sigma^{\pr}))
\pa_{\sigma^{\pr}} T(\sigma^{\pr}) \nonumber \\
&& \hskip0.5cm - {k^2 \o 6}\d (\sigma -
\sigma^{\pr}) \pa ^3 _{\sigma ^{\pr} }T(\sigma^{\pr}) 
-  {5k^2 \o 6 } ( \pa^3_{\sigma^{\pr}}\d (\sigma - \sigma^{\pr})) 
T (\sigma^{\pr}) +
{k^4 \o 6}
\pa_{\sigma^{\pr}}^5 \d(\sigma - \sigma^{\pr})
\label{14}
\er

The method we have used in the derivation of the $V^{(1,1)}_{n+1}$-algebras
($n=1,2,3$) allows us to find the corresponding fields ($\psi, \chi,
\phi_1$)-transformations.  The conformal transformations have the form :
\br
\d_{\eps }\phi_i &=& {i(i-n)\o 2}\pa \eps + \eps \pa \phi_i, \quad 
\d_{\bar \eps }\phi_i = 
{i(i-n)\o 2}\bar \pa \bar \eps + \bar \eps \bar \pa \phi_i \nonumber \\
\d_{\epsilon}\psi & =&  {(1-n) \o 2} (\pa \epsilon) \psi + \epsilon \pa \psi, \quad  
\d_{\bar \epsilon}\psi  = \bar \epsilon \bar \pa \psi \nonumber \\
\d_{\epsilon} \chi & = & \epsilon \pa \chi, \quad \quad \quad \quad \quad
\d_{\bar \epsilon} \chi  =  {(1-n) \o 2} (\bar \pa \bar \epsilon)\chi
 + \bar \epsilon \bar \pa \chi
\label{15}
\er
$i=1, \cdots ,n-1$.  The transformation generated by $V^+$  are quite
simple but highly nonlocal:
\br
\{V^+(\sigma ) ,\psi (\sigma^{\pr} )\}&=& -{{n+1}\o 4nk}\eps (\sigma 
-\sigma^{\pr})V^+(\sigma )\psi (\sigma^{\pr} ) + {1\o k}e^{{{n+1}\o 2n}
R(\sigma^{\pr} )} \d (\sigma -\sigma^{\pr} ) \nonumber \\
\{V^+(\sigma ) ,\chi (\sigma^{\pr} )\}&=& {{n+1}\o 4nk}\eps (\sigma 
-\sigma^{\pr})V^+(\sigma )\chi (\sigma^{\pr} ); \quad \{V^+(\sigma ) ,
\phi_i( \sigma^{\pr} )\} =0
\label{16}
\er
For $n>1$, the corresponding $V^-$-transformation are much more complicated
than the $V^+$-ones.

The complete algebra  of the symmetries of the NA-Toda models (\ref{1})
contains together with the two (chiral ) $V^{(1,1)}_{n+1}$ and (antichiral )
$\bar V^{(1,1)}_{n+1}$ (spanned  by $\bar V^{\pm}, \bar W_{n-i+2}$) algebras,
the PB's of the global charge $Q_0 = \int J_0 d\sigma $ of the {\it nonchiral }
$U(1)$ current:
\br
\{ Q_0 ,V^{\pm}(\sigma )\} &=& \pm V^{\pm}(\sigma ), \quad \quad 
\{ Q_0 ,\bar V^{\pm}(\sigma )\} = \pm \bar V^{\pm}(\sigma )\nonumber \\
\{Q_0, W_{n-i+2}(\sigma )\} &=& \{Q_0, \bar W_{n-i+2}(\sigma )\}= 0
\label{17}
\er
as well.  To describe the full structure of this larger algebra one has to
calculate the PB's of the chiral with antichiral currents.  As usual, the
chiral and antichiral charges of the {\it local } currents {\it do commute }:
$$
\{L_{m_1}^{(n-i+2)}, \bar L_{m_2}^{(n-j+2)}\} = 0, 
\nonumber
$$
where $L_{m}^{(n-i+2)}= \int d\sigma W_{(n-i+2)}\sigma ^{m+n-i+1}$ ( and the
same for $\bar L $).  The new phenomena occurs with the PB's of certain
nonlocal charges.  Our {\it main observation } is that $Q^+ = \int d\sigma
V^+(\sigma )$ and $\bar Q^- = \int d\sigma \bar V^-(\sigma )$ have {\it
nonvanishing } PB's:
\be
\{Q^+, \bar Q^- \} = {{k\pi }\o 2}\int_{-\infty}^{\infty} d\sigma \pa
_{\sigma}e^{{{n+1}\o n}R -\phi_1 -ln \Delta }
\label{18}
\ee
In order to prove eqn. (\ref{18}) we realize $V^+$ and $\bar V^-$ (given by
eqn. (\ref{5})) in terms of fields $\psi, \chi, \phi_i$, their space
($\pa_{\sigma }$) derivatives and their conjugate momenta $\Pi_{\psi },
\Pi_{\chi },\Pi_{\phi_i }$.  The rest is straightforward calculation based on
the canonical PB's $\{ \rho_i (\sigma ), \Pi_{\rho_j}(\sigma^{\pr}) \}
=\d_{ij}\d (\sigma - \sigma^{\pr} )$ where $\rho_i = (\psi, \chi, \phi_i )$. 
Note that the field in the exponent of thr r.h.s. of (\ref{18}):
\be
\varphi = R -{{n\o {n+1}}}(\phi_1 + ln \Delta )
\label{19}
\ee
is related to the $U(1)$-current $J_{\mu} = {k\o 2\pi }\eps_{\mu \nu}
\pa^{\nu}(\varphi + {n\o {n+1}}\phi_1 )$.  
Since the topological conserved current $I_{\mu} = {{k\o {2\pi }}}{{n\o {n+1}}}
\eps_{\mu \nu }\pa ^{\nu }\phi_1 $ has vanishing PB's with $V^{\pm}$ and $\bar
V^{\pm }$ one can redefine the $U(1)$-charge as follows:
$$
H_1 = Q_0 - \int I_0 d \sigma  = 
-{k \o {2\pi}} (\varphi (\infty) - \varphi (-\infty) ),
\nonumber
$$   
\be
\{ H_1, Q^+ \} = Q^+, \quad \quad  \{ H_1 , \bar Q^-\} = - \bar Q^-
\label{20}
\ee
Then eqn. (\ref{18}) takes the following suggestive form:
$$
\{ Q^+, \bar Q^- \} = q_{(n)}^{\d }(q_{(n)}^{H_1 }- q_{(n)}^{-H_1 })
\nonumber
$$
where $ q_{(n)} = e^{-{{n+1} \o n} ({\pi \o k})} $ and  
$\d = -{k \o {2\pi}}( \varphi
(\infty) + \varphi (-\infty))$. As a consequence of the PB's of
$\varphi(\sigma )$ with $V^+$ and $\bar V^-$ we find  that $\{\d ,Q^+ \}=
\{\d ,\bar Q^- \} =0$.  Finally introducing new charges $E_1, F_1 $ (instead
of $Q^+$ and $\bar Q^-$ ):
$$ 
E_1 = \sqrt {2 \o {k\pi}} { q^{{1+\d} \o 2} \o { (q^2 -1)}^{1 \o 2}} Q^+, 
\quad \quad  F_1 = 
\sqrt {2 \o {k\pi}} { q^{{1+\d} \o 2} \o { (q^2 -1)}^{1 \o 2}} \bar Q^- 
\nonumber
$$
we realize that the PB's algebra (\ref{18}), (\ref{20}) of $Q^+, \bar Q^-$
and $Q_0$ takes the standard form of the $q$-deformed $SL(2,R)$-algebra
(n-arbitrary ):
\be
\{E_1, F_1\} = {{q^{H_1}_{(n)} - q^{-H_1}_{(n)}} \o {q_{(n)} - q^{-1}_{(n)}}}, 
\quad \quad \{H_1, E_1 \}= E_1
,\quad \quad
\{H_1, F_1 \}= - F_1.
\label{21}
\ee
With the explicit form (\ref{7}),(\ref{8}) 
of $V^-$ and $\bar V^+$ (for $n=1,2$ only ) at hand we
can calculate  the PB's  of their charges
$$
Q^- = \int \sigma ^{n-1}V^-(\sigma ) d \sigma, 
\quad \quad \bar Q^- = \int \sigma ^{n-1}\bar V^-(\sigma ) d \sigma
\nonumber
$$
as well as the mixed PB's $\{Q^{\pm }, \bar Q^{\pm } \}$.  The result is:
\be
\{E_0, F_0\} = {{q^{H_0}_{(n)} - q^{-H_0}_{(n)}} \o {q_{(n)} - q^{-1}_{(n)}}}, 
\quad \quad \quad    \{Q^{\pm} ,\bar Q^{\pm }\} =0,\quad n=1,2
\label{22}
\ee
where(we are omiting the index (n) in q) 
$$ 
E_0 = \sqrt {2 \o {k\pi}} { q^{{1-\d} \o 2} \o {(1 - q^2)}^{1 \o 2}} Q^-, 
\quad \quad  F_0 = 
\sqrt {2 \o {k\pi}} { q^{{1-\d} \o 2} \o {( 1 - q^2) }^{1 \o 2}} \bar Q^+, 
\quad \quad \quad H_0=-H_1
\nonumber
$$
The PB's (\ref{21}) and (\ref{22}) can be written in a compact form :
\br
&&\{H_i, E_j\}  =  \kappa_{ij}E_j \quad \quad \{H_i, F_j\} = - \kappa_{ij}F_j, 
\quad \quad  i, j = 0, 1
\nonumber \\
&&\{E_i, F_j\}  =  \d_{ij}{{{q^{H_i}_n - q^{-H_i}_n} \o {q_n - q^{-1}_n}}  } , 
~~~~\kappa_{ij} =
\( \begin{array}{cc}
  1 & -1 \\
 -1 & 1
\end{array} \)
\label{23}
\er
which is known to be the {\it centerless  affine } $SL(2,R)_q$ PB's algebra
in the principal gradation \cite{Bernard} and \cite{Drinfeld} 
(the Serre relations are
omited).  The {\it conclusion } is that the {\it classical $q$-deformed
affine (for $n=1,2$) $SL(2,R)$ PB's algebra (\ref{23}), generated by certain
nonlocal charges appears as the algebra of the Noether symmetries of the
NA-Toda models (\ref{1})}.   Note that the deformation parameter $q_n =
e^{-{{n+1}\o n}({\pi \o k})}$ is a function of the classical coupling
constant $k$ and of the rank $n$ of the underlying $A_n$-algebra .

To make the above statement complete we have to demonstrate that the field
equations of (\ref{1}):
\br
\pa \bar \pa \phi_i &=& ({2\o k})e^{\phi_{i+1}+\phi_{i-1}-2\phi_i} -
{(n-i)\o n}e^{-\phi_1}{{\pa \chi \bar \pa \psi }\o {\Delta ^2}}
\nonumber \\
\bar \pa ({\pa \chi \o \Delta }e^{-\phi_1}) &=& 
-{{n+1}\o 2n}{{\bar \pa \psi \pa \chi }\o
{\Delta^2}}\chi e^{-2\phi_1}, \quad \quad 
 \pa ({\bar \pa \psi \o \Delta }e^{-\phi_1}) = -{{n+1}\o 2n}
 {{\bar \pa \psi \pa \chi }\o
{\Delta^2}}\psi e^{-2\phi_1} 
\label{24}
\er
admit solutions such that $\varphi (\infty ,t_0) \neq \varphi (-\infty
,t_0)$, i.e. with nontrivial $U(1)$-charge  $H_1 \neq 0$.  If no such
solutions exist, (i.e. $\varphi (\infty ,t_0) = \varphi (-\infty
,t_0)$)  then all PB's of the chiral and antichiral nonlocal
charges {\it vanish  identically}. 

Our construction of the solutions of eqns. (\ref{24}) is based on the
following {\it observation}: the change of the field variables $\phi_i, \psi ,
\chi $ into $\varphi_i ,V^+, \bar V^- $ given by 
\br
\varphi_i & =&  \phi_{i-1} + 
{{n-i+1}\o n}R + (i-1)\ln(V^+\bar V^-), \quad \phi_0 =0 \nonumber \\
\psi V^+ &=& ({k\o 2})e^{{{n+1}\o n}\varphi_1}\pa \varphi_1, \quad 
 \chi \bar V^- = ({k\o 2})e^{{{n+1}\o n}\varphi_1}\bar \pa \varphi_1
\label{25}
\er
(the last two equations reflect the definition (\ref{5}) of $V^+, \bar V^-$
 and (\ref{6}) of $R$) {\it maps} the eqns. (\ref{24}) into the following
system of equations :
\br
\pa \bar \pa \varphi_l &=& ({2 \o k})^2 e^{\varphi_{l-1} + \varphi_{l+1}
 -2\varphi_l},
\quad l=1, \cdots ,n-1 \nonumber \\
\pa \bar \pa \varphi_n &=& (V^+ \bar V^- )^n e^{-2\varphi_{n} + \varphi_{n-1}},
\quad \pa \bar V^- =\bar \pa V^+ = 0 
\label{26}
\er 
The general solution of eqns. (\ref{26}) can be found by slight
modification of the Gervais-Bilal method \cite{Bilal1}, realizing the
$\varphi_i, (i=1, \cdots ,n)$ and $V^+, \bar V^-$ in terms of
$n+1$-independent functions $f_l(t+ \sigma ), \bar f_l(t -\sigma ), l=1,2,
\cdots ,n+1$:
$$
e^{\varphi_1} = ({k \o 2})^{-n}f_l\bar f_l, \quad \cdots \quad 
e^{\varphi_p} = ({1\o 2})^{p(p-n-1)}{1\o {p!}}f_{l_1, \cdots l_p} 
\bar f_{l_1, \cdots l_p}
\nonumber 
$$
\be
(V^+)^n =  \epsilon_{l_1,l_2, \cdots l_{n+1}}f_{l_1}f^{(1)}_{l_2}
 \cdots f^{(n)}_{l_{n+1}}; \quad (\bar V^+)^n =  
 \epsilon_{l_1,l_2, \cdots l_{n+1}}\bar f_{l_1}\bar f^{(1)}_{l_2}
 \cdots \bar f^{(n)}_{l_{n+1}}
 \label{27}
\ee
where $f_{l_1, \cdots l_p} $ are rank p antisymmetric tensors.  For example,
 $f_{l_1, l_2} = f_{l_1}f^{\pr}_{l_2}- f^{\pr}_{l_1}f_{l_2}$.  Then the
 solution of eqns. (\ref{24}) is given by:
$$
e^R = ({k\o 2})^{-n}f_l\bar f_l , \quad \quad e^{\phi_1} = 
{1\o 2}({k\o 2})^{1-n} (f_{lm}\bar f_{lm})(f_p\bar
f_p)^{{{1-n}\o n}}(V^+\bar V^-)^{-1}, etc. 
\nonumber
$$ 
\be
\psi = ({k \o 2})^{{1-n}\o 2} 
(f_l\bar f_l)^{{{1-n}\o 2n}}(f_p^{\pr}\bar f_p) (V^+)^{-1},
 \quad \quad 
\chi = ({k \o 2})^{{1-n}\o 2} 
(f_l\bar f_l)^{{{1-n}\o 2n}}(f_p\bar f_p^{\pr}) (\bar
V^-)^{-1}
\label{28}
\ee
Therefore the field $\varphi $ from eqn. (\ref{19}) whose assymptotics are
under investigation takes the form:
\be
\varphi = -{n \o{n+1}} ln ({k\o 2})^2\( {{(f_l^{\pr} \bar f_l^{\pr})(f_p\bar f_p) - 
{{n-1}\o 2n}
(f_l^{\pr}\bar f_l)(f_p\bar f_p^{\pr})} \o {(f_m \bar f_m )^2 
(V^+\bar V^-)}}\) 
\label{29}
\ee
We next consider the following ansatz
\be
f_l = \a_l e^{{(t+\sigma )}a_l}, \quad \quad 
\bar f_l = \bar \a_l e^{{(t-\sigma )}\bar a_l},\quad\quad
 \sum_{l=1}^{n+1} (a_l - \bar a_l ) = 0 
\label{30}
\ee
and for convenience we choose the following parametrization for the $a_l$'s:
\be
a_1 - \bar a_1 = b_1 + b_2 + \cdots + b_n; \quad \quad a_p - \bar a_p =
 -b_{p-1}, \quad
b_n >b_{n-1}> \ldots > b_1
\nonumber
\ee
where $p=2,3, \ldots , n+1 $.
Under all these conditions we calculate the limits $\sigma \longrightarrow
\pm \infty $ of eqn. (\ref{29}) at $t=0$:
$$ 
\varphi (\infty ,0) = -{n \o {n+1}}\ln (({k\o 2})^2{{n+1}\o 2n}
 a_1 \bar a_1 A), 
 \quad 
\varphi (-\infty ,0) = -{n \o {n+1}}\ln (({k\o 2})^2{{n+1}\o 2n} a_{n+1}
 \bar a_{n+1} A), 
 \nonumber
$$
$A=V^+\bar V^-(t=0)$.
Since $a_1\bar a_1 \neq a_{n+1}\bar a_{n+1}$ the solutions (\ref{30}) are an
example of solutions of (\ref{24}) with $H_1 \neq 0$.

The quantization of the classical $V^{(1,1)}_{n+1}$-algebras represents
certain new features all related to the nonlocal terms $\eps (\sigma )$ in the
r.h.s. of (\ref{11}).  We find more convenient to directly apply the procedure
of {\it quantum } Hamiltonian reduction to the $A_n$-WZW models instead of
quantizing the results of the classical Hamiltonian reduction.  The method we
are going to use is an appropriate generalization  of the derivation of the
parafermionic algebra \cite{Zamolodchikov} from the affine $SU(2)$-one 
(or $SL(2,R)$
for the noncompact PF's) by imposing the constraint $J_3 =0$.  Following the arguments of
ref. \cite{Zamolodchikov} we define the quantum (compact )$V_2$-algebra as
$$
V_2 = \{SU(2)_k , J_3 =0 \}
\nonumber 
$$
Therefore its generators $\psi^{\pm}$ represent the $J_3 = \sqrt {{k\o 2}}\pa
\phi $ independent part of the $\hat SU(2)_k$-ones, namely, $J^{\pm}$, 
\br
J^{\pm} = \psi^{\pm}e^{\mp \a \phi }, & T = T_V + {1\o 2}(\pa \phi )^2
\nonumber \\
J_3(z_1) \psi^{\pm }(z_2) = O(z_{12}), & \phi(z_1) \phi(z_2) = 
-ln z_{12} + O(z_{12})
\label{31}
\er
Taking into account the $SU(2)$ OPE's:
$$
J_3(z_1) J^{\pm}(z_2) = \pm {i \o z_{12}}J^{\pm}(z_2) + O(z_{12})
\nonumber
$$
and eqns. (\ref{31}) we find $\a = i\sqrt {{2\o k}}$.  Another consequence of
eqn. (\ref{31}) is that the dimensions of $\psi^{\pm}$ are $\Delta^{\pm} = 1-
{1\o k}$ ( we have used that $\Delta_{J^{\pm}}=1$).  Finally the construction (\ref{31}), the $ \phi(z_1) \phi(z_2) $ --OPE
and the $SU(2))_k $--OPE's leads to  the  following $V_2$-algebra OPE's:
\br
\psi^{\pm}(z_1)\psi^{\pm}(z_2) &=& z_{12}^{-{2\o k}}\psi^{\pm}_{(2)}(z_2) + O(z_{12})
\nonumber \\
\psi^{+}(z_1)\psi^{-}(z_2) &=& z_{12}^{{2\o k}}({k\o {z_{12}^2}} + (k+2)T_V(z_2)
+ O(z_{12})),
\label{32}
\er
which is nothing but the  well known PF-algebra \cite{Zamolodchikov}.  Although the
PF ($V_2$)-algebra (\ref{32}) is {\it by construction } the quantum version of
the classical PB's algebra (see eqn. (\ref{11}) for $n=1$):
\br
\{ V^{\pm}(\sigma ), V^{\pm}(\sigma^{\pr})\} &=&
 -\eps (\sigma-\sigma^{\pr}) V^{\pm}(\sigma ) V^{\pm}(\sigma^{\pr})\nonumber
 \\
\{ V^{-}(\sigma ), V^{+}(\sigma^{\pr})\} &=& 
\eps (\sigma-\sigma^{\pr}) V^{+}(\sigma ) V^{-}(\sigma^{\pr}) +
\pa_{\sigma^{\pr}} \d (\sigma -\sigma^{\pr})
\label{33}
\er
the discrepancy between the dimensions $\Delta^{\pm} =1-{1\o k}$ of
$\psi^{\pm}$ and $\Delta^{\pm}_V =1$ of $V^{\pm}$ requires more precise
definition of the relation  of eqn. (\ref{32}) and (\ref{33}).  The exact {\it
statement } is as follows:
Let $V^{\pm} = {1\o k}\psi^{\pm} $ and the $V^{\pm}$ PB's are defined as
certain limit of the OPE's (\ref{32}):
\be
\{ V^a(z_1), V^b(z_2) \} = {\rm lim}_{k \rightarrow \infty }
 {k\o {2\pi i}}( V^a(z_1) V^b(z_2)-  V^b(z_2)V^a(z_1))
 \label{34}
 \ee
( $a,b = \pm $).  Then the $k\longrightarrow \infty $ limit of the OPE's
(\ref{32}) reproduces the PB's algebra  (\ref{33}).  The proof is
straightforward.  Applying twice the OPE's (\ref{32}) we obtain 
$$
z_{12}^{2\o k} (V^{\pm}(z_1)V^{\pm}(z_2)-e^{-{{2\pi i}\o k}\eps (z_{12})}
V^{\pm}(z_2)V^{\pm}(z_1)) = {1\o k^2}O(z_{12}) \nonumber 
$$
\be
z_{12}^{-2\o k} (V^{-}(z_1)V^{+}(z_2)-e^{{{2\pi i}\o k}\eps (z_{12})}
V^{+}(z_2)V^{-}(z_1)) = {1\o k} ({1\o {z_{12}^2+i0}} - {1\o {z_{21}^2+i0}}) +
{{k+2}\o k^2}O(z_{12})
\label{35}
\ee
where the identity $i\pi \eps (z_{12}) = ln {{z_{12}+i0}\o {z_{21}+i0}}$ has
been used.  The $k \rightarrow \infty$ limit of (\ref{35}) reproduces the
PB's (\ref{33}) of the classical $V_2$-algebra \footnote {The noncompact case
$SL(2,R)/U(1)$ corresponds to the change $\phi \longrightarrow i\phi $, which
turns out to be equivalent to $k\rightarrow -k $ in the OPE's, spins,
etc.}.   The conclusion is that the nonlocal PB's algebra (\ref{33}) is {\it
semiclassical limit } ($k \rightarrow  \infty $) of the PF-algebra and
that {\it the quantization} of the nonlocal currents  $V^{\pm}$ requires
{\it renormalization } of their (classical ) spins: $\Delta^{\pm}_q = \Delta^{\pm}_{cl} - {1\o k}$.
For $k$-positive integers the global $Z_2\otimes U(1)$ symmetry of the
classical $A_1$-NA-Toda model is broken to the $Z_2\otimes Z_k $ of the
quantum theory.

The quantization of the $V_3^{(1,1)}$-algebra is based on the following
{\it observation }:  the $A_2^{(1,1)}$-NA-Toda model is equivalent to the
$U(1)$-{\it reduced } Bershadsky-Polyakov $A_2^{(2)}$-NA-Toda  model (BP)
\cite{Bershadsky2,Polyakov}.  In fact, the set of constraints (and
gauge fixing conditions ) (\ref{2}), (\ref{3}) for $n=2$ appears to be the
image of the BP-ones \cite{Bershadsky2}
\be
J_{-\a_2} = \bar J_{\a_2} = 0, \quad  J_{-\a_1-\a_2} = \bar J_{\a_1+\a_2} =1
\label{36}
\ee
($J_{-\a_1} =0$ is the gauge fixing condition for the constraint$ J_{-\a_2}=0$)
and the additional constraint
\be
J_{(\lambda_1 -\lambda_2 )\cdot H} = \bar J_{(\lambda_1 -\lambda_2 )\cdot H}=
0
\label{37}
\ee
under specific Weyl reflection $\omega_{\a_1}(\a ) = \a_1 - (\a \cdot \a_1 )\a
$.  The constraint (\ref{37}) imposed on the $U(1)$ current transforms
$W_3^{(2)}$-algebra (the symmetry of the original BP-model ) into the nonlocal
algebra $V_3^{(2)} \equiv V_3^{(1,1)}$ :
\be
V_3^{(2)} = \{ W_3^{(2)} ; J_{(\lambda_1 -\lambda_2 )\cdot H} = 0\}
\label{38}
\ee
The {\it statement } is that $A_2^{(2)}$ and $A_2^{(1,1)}$-models have
identical algebras of symmetries $V_3^{(2)} = V_3^{(1,1)}$ (see eqns.
(\ref{11})) and their Lagrangeans :
\br
{\cal L}_2^{(2)} &=& -{k\o 2\pi }
\left( \pa \varphi_0 \bar \pa \varphi_0 +
{{e^{\varphi_0}\bar \pa \psi_0 \pa \chi_0 }\o {1+ {3\o 4}e^{\varphi_0}\psi_0
\chi_0 }} - e^{-2\varphi_0} (1+ \psi_0 \chi_0 e^{\varphi_0})\right) \nonumber \\
{\cal L}_2^{(1,1)} &=& -{k\o 2\pi }\left(\pa \varphi \bar \pa \varphi +{{e^{-\varphi}\bar \pa \psi \pa \chi }\o {1+ {3\o 4}e^{-\varphi}\psi
\chi }} - e^{-2\varphi} \right)
\nonumber
\er
are related by the following change of the variables:
 \br
\psi = \chi_0 e^{\varphi_0} (1+  e^{\varphi_0}\psi_0 \chi_0 )^{-{1\o 4}},\,\,
 \chi = \psi_0 e^{\varphi_0} (1+  e^{\varphi_0}\psi_0 \chi_0 )^{-{1\o 4}},\,\,
 \varphi = \varphi_0 - {1\o 2} ln (1+  e^{\varphi_0}\psi_0 \chi_0 )
\nonumber
\er
 i.e. ${\cal L}_2^{(2)} = {\cal L}_2^{(1,1)}$ + total derivative. \footnote
 {The detailed proof is present in our forthcoming paper \cite{GSZ}}  This fact,
 together with the OPE's of the $W_3^{(2)}$-currents $G^{\pm}, T_W, J 
 (\equiv J_{(\lambda_1 -\lambda_2 )\cdot H} )$ ( of spins $\Delta_{G^{\pm}} =
 {3\o 2}, \Delta_T =2, \Delta _J = 1$) (see ref. \cite{Bershadsky2})
 \br 
&& J(z_1) G^{\pm}(z_2) = \pm {1\o z_{12}} G^{\pm}(z_2) + O(z_{12});\,\,\, J(z_1)J(z_2)
 = {(2k+3)\o 3z_{12}}+ O(z_{12}), \nonumber \\
 &&G^{\pm}(z_1)G^{\pm}(z_2)=O(z_{12}),\, {\rm etc.},
 \label{39}
 \er
lead us to the following relation between $G^{\pm}, T_W$ , $J= \sqrt
{{(2k+3)\o 3}}\pa \tilde \phi$ and the $V_3^{(1,1)}$-currents $V^{\pm}, T_V$:
 \be
 G^{\pm} = V^{\pm} e^{\pm \sqrt {{3\o {2k+3}}}\tilde \phi }, \quad T_W = T_V +
 {1\o 2} (\pa \tilde \phi )^2.
 \label{40}
 \ee
 Remind that according to (\ref{38}) we have to impose 
 \be J(z_1) V^{\pm}(z_2)=O(z_{12}) = J(z_1) T_{V}(z_2).
 \label{41}
 \ee
  and that the OPE's (\ref{39}) are compatible with the bosonization  of the
  $U(1)$ current $J$ if the following OPE 
  \be
  \tilde \phi (z_1)\tilde \phi (z_2) = \ln z_{12} + O(z_{12})
  \label{42}
  \ee
  takes place.  As a consequence of eqns. (\ref{39}), (\ref{40}), (\ref{41})
  and (\ref{42}) we find that the spins of the {\it quantum currents } $V^{\pm}$
  are {\it renormalized } ($\Delta^{\pm}_{cl} = {3\o 2}$)
  $$
  \Delta ^{\pm}_q = {3\o 2} - {3\o {2(2k+3)}}
  \nonumber 
  $$
  and that the $V^{\pm}$ and $T_V$-OPE's (that define the quantum $V_3^{(1,1)}$) have the form ($k=-3,-{3\o 2}, -1$):
  \br
  V^{\pm}(z_1)V^{\pm}(z_2)&=& z_{12}^{-{3\o {2k+3}}}V_{(2)}^{\pm}(z_2) + O(z_{12})
  \nonumber \\
   V^{+}(z_1)V^{-}(z_2)&=&  z_{12}^{{3\o {2k+3}}}
\left( {{(2k+3)(k+1)}\o {z_{12}^{3}}} -
   {{k+3}\o z_{12} }T_V(z_2)\right) +  O(z_{12})\nonumber \\
   T_V(z_1) V^{\pm}(z_2)&=& {\Delta^{\pm} \o z_{12}^{2}} V^{\pm }(z_2) + {1\o
   z_{12}} \pa V^{\pm}(z_2)+  O(z_{12})
   \label{43}
   \er
The $T_V(z_1)   T_V(z_2)$ has the standard form of the Virasoro algebra OPE
 with central charge $c_V = c_W - 1 = -6 {(k+1)^2 \o (k+3)}$.   The
 $V_3^{(1,1)}$-algebra (\ref{43}) is quite similar to the standard PF-algebra
 \cite{Zamolodchikov} and for $L$-positive integers ($L>3)$ the OPE's (\ref{43}) involve
 more currents $V_l^{\pm}$, $l=1,2, \cdots ,L-1$, of dimensions $\Delta^{\pm}_l
 = {3\o 2} {l\o L} (L-l), L=2k+3$.  Following the arguments of ref.
 \cite{Zamolodchikov} we
 define  (Laurent ) mode expansion for the currents   $V^{\pm}$ :
 \be
 V^{\pm}(z) \phi_s^{\eta }(0) = \sum_{m=-\infty }^{\infty} z^{\pm {3s \o 2L} +
 m -1 \mp \eta }V^{\pm}_{-m\pm \eta - {1\o 2} + {3(1\mp s)\o 2L}} 
  \phi_s^{\eta }(0)
  \label{44}
  \ee
 where $ \phi_s^{\eta }(0)$ denote certain Ramond ($\eta =1/2, s -odd$) and
 Neveu-Schwarz ($\eta =0, s -even$) fields, $s=1,2, \cdots L-1$.  Then the
 OPE's (\ref{43}) give rise to the following {\it PF-type} 
$\quad \quad \quad \quad \quad$ {\it `` commutation
 relations''} for the $V_3^{(1,1)}(L)$-algebra ($|L|>3$)
 \br
&&\hskip -0.5cm {2\o {L+3}}\sum_{p=0}^{\infty }C_{(-{3\o L})}^p \left(V^{+}_{-{{3(s+1)}\o
{2L}}+m-p-\eta +{1\o 2}}V^{-}_{{{3(s+1)}\o {2L}}+n+p+\eta -{1\o 2}}
+V^{-}_{-{{3(1-s)}\o {2L}}+n-p+\eta -{1\o 2}}V^{+}_{{{3(1-s)}\o
{2L}}+m+p-\eta +{1\o 2}}\right) \nonumber\\
&&= -L_{m+n} + {{(L-1)L}\o {2(L+3)}}({{3s}\o {2L}}+n+\eta )
({{3s}\o {2L}}+n+\eta -1)\d_{m+n,0}
\label{45}
\er
where $C^p_{(M)} = {{\Gamma (p-M)}\o {p!\Gamma (-M)}}$, 
 $m,n =0, \pm 1, \pm 2,\cdots $ and 
  \be
\sum_{p=0}^{\infty }C_{({3\o L})}^p 
\left(V^{\pm}_{{{3(3\mp s)}\o
{2L}}-p+m+\eta -{1\o 2}}V^{\pm}_{{{3(1\mp s)}\o {2L}}+p+n+\eta -{1\o 2}}-
V^{\pm}_{{{3(3\mp s)}\o
{2L}}-p+n+\eta -{1\o 2}}V^{\pm}_{{{3(1\mp s)}\o {2L}}+p+m+\eta -{1\o 2}}\right)=0
\label{46}
\ee
 In the particular cases $L=2,3$ the OPE's $V^{\pm}V^{\pm}$ have also a pole ,
 which makes eqn. (\ref{46}){\it nonvalid}.  The simplest example of such $V_3^{(1,1)}$ 
algebra $L=2$ is spanned by $V^{\pm}$ of $\Delta^{\pm}_{L} = {3\o 4}$ and $T_V$
only.  Its central charge is $c_V(L=2) = -{3\o 5}$.  The relations (\ref{46}) are
now substituted by:
 $$
\sum_{p=0}^{\infty }C_{({1\o 2})}^p \left(V^{-}_{-p+m+\eta -{3\o 4}}
V^{-}_{p+n+\eta -{5\o 4}}+
V^{-}_{-p+n+\eta -{3\o 4}}V^{-}_{p+m+\eta -{5\o 4}}\right)= \d_{m+n+2\eta ,0}
\nonumber
$$ 
and similar one for $V^+V^+$'s.  Again as in the $n=1$ case one can verify that
certain limit of the OPE's (\ref{43}) reproduces the classical PB's
$V_3^{(1,1)}$-algebra (\ref{11}).

The relation (\ref{40}) between $W_3^{(2)}$  and  $V_3^{(2)}$ currents leads to
the following form of the $W_3^{(2)}$-(chiral) vertex operators
$\phi^{W}_{(r_i,s_i)}(z)$, $i=1,2$ in terms of the $V_3^{(1,1)}$-ones 
$\phi^{V}_{(r_i,s_i)}$  and $\tilde \phi $:
\be
\phi^{W}_{(r_i,s_i)} = \phi^{V}_{(r_i,s_i)} exp (q_{(r_i,s_i)}\sqrt {{3\o
L}}\tilde \phi )
\label{47}
\ee
The construction (\ref{47}) is a consequence of (\ref{40}), (\ref{42}), the
OPE's
\br
T^{W}(z_1)\phi ^{W}_{(r,s)}(z_2) & = &
{{\Delta^{W}_{r,s}}\o {z_{12}^2}}\phi ^{W}_{(r,s)}(z_2)  +
 {1\o {z_{12}}}\pa \phi ^{W}_{(r,s)}(z_2) + O(z_{12})
 \nonumber\\
J(z_1)\phi ^{W}_{(r,s)}(z_2) & = &{{q_{r,s}}\o {z_{12}}}
 \phi^{W}_{(r,s)}(z_2) + O(z_{12}), 
 \nonumber 
\er
which define $ \phi^{W}_{(r,s)}$ as $W_3^{(2)}$ primary fields and the
 fact that
$\phi_{(r,s)}^V $ are $J$-neutral, i.e. $J(z_1) \phi_{(r,s)}^V(z_2) = 
O(z_{12})$.  Finally we realize that the dimensions  of the
$V_3^{(1,1)}$-primary fields $\phi_{(r,s)}^V $ are related to the 
$\phi_{(r,s)}^W $-dimensions and charges by the following formula:
\be 
\Delta^V_{(r,s)} = \Delta^W_{(r,s)}- {3\o 2L}q_{(r,s)}^2
\label{48}
\ee
Taking into
account  the explicit values of $\Delta^W_{(r,s)}$ and $q_{(r,s)}$ for the
class of `` completely  degenerate'' highest weight representations of
$W_3^{(2)}$ (which for rational levels $L+3={4p\o q}$ have been calculated in
ref. \cite{Bershadsky2})  we find that the conformal dimensions 
$\Delta^V_{(r,s)}$ of the `` degenerate'' representations of $V_3^{(1,1)}$ are
given by :
\br
&&\hskip -0.7cm \Delta_{(r,s)}^{V} = {{1\o {32(L+3)}}}\left((L-3)((L+3)r_{12}-4s_{12})^2
%\nonumber
%\\
+4L((L+3)r_1-4s_1)((L+3)r_2-4s_2)\right)- \nonumber\\
&&\hskip 1cm -{4L(L-1)^2\o 32(L+3)} -{{\eta^{W}}\o {8L}}[L+3\eta^{W} \pm ((L+3)r_{12}-4s_{12})],
\label{49}
\er
$1\leq r_i \leq 2p-1$, $1\leq s_i \leq 2q-1$, $r_{12} = r_1 -r_2$ where
$\eta^{W} =0$, $r_i$-odd integers for  the NS-sector,
 $\eta^{W} =1/2 $, $r_i$-even integers for  the Ramond-sector and $\eta^W = {1\o
 2} - \eta$. 

 The parafermionic features of the $V_2$ and $V_3^{(1,1)}$-algebras  rises the
 question 
 whether the quantum $V_{n+1}^{(1,1)}$ -algebras 
 share these properties.  Our preliminary calculations of the {\it renormalized
 } spins of the {\it nonlocal } currents $V^{\pm}_{(n)}$ (for the
 $A_n^{(1,1)}$-model )
 $$
 \Delta^{\pm}_n = {{n+1}\o 2}(1 - {1\o {2k+n+1}})
 \nonumber
 $$
 shows that this is indeed the case.  An interesting open question is about the
 {\it quantum counterpart } of the classical $\hat SL(2,R)_q$-PB's algebra
 (\ref{21}), (\ref{22}), (\ref{23}).  Although we have no satisfactory answer to
 this question, the particular case $n=1$, $k=2$ (critical Ising model)
 provides a promissing hint.  The quantum nonlocal charges $Q^+$ and $\bar Q^-$
 coincide in this case with the Ramond sector's zero modes $\psi_0$, $\bar
 \psi_0$ of the Ising fermions.  Due to the double degeneracy of the lowest
 energy state $| \sigma _{\pm} \rangle $
 $$ 
\psi_0 | \sigma_{\pm}> = {1 \o {\sqrt {2}}}| \sigma_{\mp}>\quad \quad \quad
\bar \psi_0 | \sigma_{\pm}> = \mp i{1 \o {\sqrt {2}}}| \sigma_{\mp}>
\nonumber
$$
their commutator does not vanish (see Sect. 6 of ref. \cite{Furlan}),
\be
[\psi_0 , \bar \psi_0]| \sigma_{\pm}> = i \Gamma | \sigma_{\pm}> \quad \quad 
\ee
where $\Gamma$ is the fermion parity operator .  The nonvanishing commutator of
the ``left'' and ``right '' fermionic zero modes {\it is not in contradiction
with the holomorphic factorization } of the critical Ising model.  What is
important is that the {\it anticommutator } $[\psi_0, \bar \psi_0 ]_+ = 0 $ 
indeed vanishes. 

{\it Acknowlegments }.One of us (GS) thanks the Department of Theoretical Physics,
UERJ-Rio de Janeiro  for the hospitality and  financial support. GS also thanks  IFT-UNESP,Laboratoire de Physique Mathematique,Universite de Montpellier II, DCP-CBPF and Fapesp 
 for the partial financial support at the initial and the final stage  of this work.
(JFG) thanks ICTP-Trieste for hospitality and support where part
 of this work
was done. This work was partially supported by CNPq.

%%%%%%%%%%%%%%%%%%%%%%%%%%%%%%%%%%%%%%%%%%%%%%%%%%%%%%%%%%%%%%%%%%%%%%%%%%%%%%%  

%%%%%%%%%%%%%%%%%%%%%%%%%%%%%%%%%%%%%%%%%%%%%%%%%%%%%%%%%%%%%

\end{document}